\documentclass{aa} 

\let\linenumbers\nolinenumbers

\usepackage{graphicx}
\usepackage{txfonts}
\usepackage{lipsum}
\usepackage{subcaption}
\usepackage{lscape}
\usepackage{placeins}

\newcommand{\lsi}{LS~I~+61~303}
\newcommand{\gammaray}{$\gamma$-ray}
\newcommand{\fermilat}{\textit{Fermi}-LAT}

\begin{document}

\nolinenumbers

\title{
  Identifying the physical periods in the radio emission from the \gammaray{} emitting binary \lsi{}
}

\titlerunning{
  Intrinsic periods in \lsi{}
}

\author{
  F. Jaron\inst{1,2}
  \and
  V. Bosch-Ramon\inst{3}
}

\institute{
  Technische Universit\"at Wien, Wiedner Hauptstra{\ss}e 8-10, 1040, Vienna, Austria\\
  \email{Frederic.Jaron@tuwien.ac.at}
  \and
  Max-Planck-Institut f\"ur Radioastronomie, Auf dem H\"ugel 69, 53121, Bonn, Germany
  \and
  Departament de F\'isica Qu\`antica i Astrof\'isica, Institut de Ci\`encies del Cosmos (ICC), Universitat de Barcelona (IEEC-UB), Mart\'i i Franqu\`es 1, E08028 Barcelona, Spain
}

\date{Received September 30, 20XX}

\abstract
{
  The $\gamma$-ray emitting binary system \lsi{} exhibits periodic emission across the electromagnetic spectrum, extending from radio frequencies up to the very-high-energy regime. The most prominent features are three periods with values $P_1 = 26.5$~d, $P_2 = 26.9$~d, and $P_{\rm long} = 4.6$~years. Occasionally,  a fourth period of 26.7~d is also detected. Mathematically, these four periods are interrelated via the interference pattern of a beating. Competing scenarios that seek to determine which of these periods are physical and which are secondary are under debate. In addition, the detection of a fifth period, $P_3 = 26.3$~d, was recently claimed. 
}
{
  Our aim is to determine which of these periods are intrinsic (i.e., whether they are likely to be related to physical processes) and which of these are secondary (i.e., resulting from interference). We chose to avoid any assumption concerning the physical scenario and restricted our analysis to the phenomenology of the radio emission variability.
} 
{
  We selected intervals from archival radio data and applied the generalized Lomb-Scargle periodogram. We fit the observational data to generate synthetic datasets that only contain specific signals in the form of pure sine waves. We analyzed these synthetic data to assess the impact of these signals and their interference on the light curves and the periodogram. 
}
{
  The two-peaked profile, consisting of $P_1$ and $P_2$, was detected in the periodogram of the actual data for intervals that are significantly shorter than $P_{\rm long}$, provided that these intervals contain the minimum of the long-term modulation. The characteristics of the observational data and their periodogram could only be reproduced with synthetic data if these explicitly included all of the three periods,~$P_1$, $P_2$, and~$P_{\rm long}$, with the residuals being limited by noise.
}
{
  We have found, for the first time, that all three periods, $P_1$, $P_2$, and $P_{\rm long}$, could, in fact, correspond to physically real processes occurring in \lsi{}. This should be considered when modeling the physical processes in this binary system.
}


\maketitle

\section{Introduction}

The system \lsi{} is unique among the class of \gammaray{} emitting binaries \citep{Dubus2013,bor24} because its emission across the electromagnetic spectrum, from the radio up to the very-high-energy regime, is characterized by the presence of several periodic signals. Timing analyses have thus far resulted  in the firm detection of four period values. The first period, $P_1$, has been identified with the orbital period of the binary system at $P_1 = 26.5$~d \citep{Taylor1982, Gregory2002}. Timing analyses of optical polarization data, presented by \citet{Kravtsov2020}, indeed corroborates that $P_1$ is the orbital period. The second period is $P_2 = 26.9$~d \citep{Massi2013, Wu2018, Jaron2024}. Competing physical scenarios are being discussed \citep[][and references therein]{Chernyakova2023, Jaron2024}, but we have decided not to presuppose any specific physical process with $P_2$ and, consequently, we refer to this period as quasi-orbital throughout this article. A long-term period was detected with a value of $P_{\rm long} = 4.6$~years (first suggested by \citealt{ParedesPhD}, then confirmed by \citealt{gre89}; \citealt{par90,Gregory2002, Massi2016, Jaron2024}); this period is also referred to as the super-orbital period of the system. The long-term period manifests itself as a modulation of the peak flux densities of radio outbursts, which occur periodically once per orbit, and in a saw-tooth-like modulation of their orbital phase occurrence \citep{Gregory2002}. As with $P_2$, we mostly discuss $P_{\rm long}$ from a phenomenological point of view and we do not presuppose any specific physical origin to it. The fourth period reported for this source is $P_{\rm average} = 26.7$~d \citep{Ray1997}, which we call ``average'' because its value is coincident with the arithmetic average of $P_1$ and $P_2$. As pointed out by \citet{Jaron2013}, this is the period with which the radio outbursts actually occur, performing a 0.5 phase jump during each minimum of the super-orbital modulation. It was first  pointed out by \citet{Massi2013} that all of these four periods ($P_1$, $P_2$, $P_{\rm long}$, and $P_{\rm average}$) are interconnected with each other via the mathematical concept of a beating. In this scenario, at least two of the four periods are intrinsic (i.e., physical) periods, being related to physical processes at work in the source, while other periods may be the result of their interference. All of these periodic signals have remained remarkably stable over the observational history of the source. In particular, the long-term modulation has by now completed ten cycles since the beginning of radio observations of \lsi{} \citep{Jaron2024}.

The presence of these periods has been confirmed for the emission at other wavelengths as well. \cite{par97} found evidence of periodicity in X-rays with a period of 26.7~d. \citet{app01} first suggested  that the X-ray emission from \lsi{} is subject to the long-term modulation, which was later confirmed by \citet{Li2012}, and \citet{Dai2016} reported the presence of~$P_1$ and~$P_2$. Concerning the giga-electron volt (GeV) \gammaray{}s, as observed by the \fermilat{}, \citet{Hadasch2012} detected the long-term modulation and \citet{Jaron2014} detected $P_1$ and $P_2$ (later confirmed by \citealt{Chernyakova2023}).
A fifth period, $P_3$, has recently been reported by \citet{Zhang2024} to be present in GeV emission. This author performed timing-analysis of \fermilat{} aperture photometry data and detected, along with the above mentioned $P_1$ and $P_2$, a third period in that range with a value of $P_3 = 26.3$~d. The presence of such a period could, however, so far not be independently confirmed. In their reference to Fig.~3~(d) of \citet{Jaron2018}, \citet{Zhang2024} put importance to a feature in the periodogram of radio data that was not identified as significant by the authors. A detailed analysis of an updated dataset by \citet{Jaron2024} did also not result in the detection of any significant feature at a period compatible with $P_3$. Concerning the GeV emission itself, \citet{Chernyakova2023} presented a periodogram of the energy range 0.1-0.3~GeV,  where, according to \citet{Zhang2024}, $P_3$ should best be detected (black curve in their Fig.~5); however, such a feature is completely absent there (see blue curve in Fig.~2 of \citealt{Chernyakova2023}). In the analysis of radio data we carried out in this work, we revisited the issue of the possible presence of such a feature and payed attention if a period compatible with $P_3$ was significantly detected or not.

The debate surrounding the actual nature of the different periods detected in \lsi{}, some of which might be the result of a beating between two of the intrinsic (i.e., physical) periods, must be resolved before a robust conclusion can be reached regarding the intrinsic physical processes behind those periods. Thus, the objective of the analysis presented here is to investigate what the intrinsic periods of the system \lsi{} are. For this purpose, we applied a timing analysis to archival radio data, selecting time intervals of these data. The intervals were chosen such that only intrinsic signals would appear in the periodogram. Furthermore, we generated synthetic data that contain only certain periodic signals to systematically analyze the effect of these signals and their interference on the light curve and the periodogram. In all of our analysis, we avoided putting any constraints that might arise from considering a certain scenario concerning the nature of the physical components or processes at work in the binary system. 

The paper is structured as follows. Section~\ref{sec:lsi} gives a brief introduction to the astrophysical system \lsi{} and a summary of the periodic signals found in its electromagnetic emission. We give a description of the datasets that we used for our analysis in Sect.~\ref{sec:data}. The methods for timing analysis are described in Sect.~\ref{sec:methods}. We present and discuss our results in Sect.~\ref{sec:results}, followed by our conclusions in Sect.~\ref{sec:conclusions}.

\section{The stellar system \lsi{}} \label{sec:lsi}

\subsection{The binary system}

The optical component in the stellar binary system \lsi{} is a B0 Ve star \citep{Hutchings1981}. The origin of optical emission lines in Be stars is usually attributed to the presence of an equatorial decretion disk \citep[][and references therein]{Rivinius2013}. The orbital parameters of the system were determined by \citet{Casares2005}, who found that the eccentricity of the orbit is $e = 0.72 \pm 0.15$. A slightly less eccentric orbital solution was found by \citealt{ara09}, who reported $e = 0.537 \pm 0.034$. The most precise determination of the orbital period is $P_1 = 26.4960 \pm 0.0028$~d by \citet{Gregory2002}, who also estimated the long-term period as $P_{\rm long} = 1667 \pm 8$~d, by the analysis of long-term radio flux monitoring data. The quasi-orbital period~$P_2$ was first detected by \citet{Massi2013} during a revisit of these radio data. This period has also been found in the X-ray emission from the source \citep{Dai2016} and in GeV \gammaray{}s \citep{Jaron2014, Chernyakova2023}. The most precise measurement of this period has been achieved with direct radio interferometric observations, carried out by \citet{Wu2018}, who made use of phase-referenced astrometry and found $P_2 = 26.926 \pm 0.005$~d. 

The second component in the binary system is a compact object, whose nature cannot be clarified from the radial velocity data and orbital solution alone. \citet{Casares2005} came to the conclusion that, depending on the inclination angle, both a neutron star and a black hole are possible. The detection of radio pulses from the direction of the system by \citet{Weng2022} certainly strengthens the case for the compact object being a neutron star, rather than a black hole. However, even if an origin in \lsi{} for these radio pulses is the simplest and more likely explanation, it would be important to increase the number of radio pulse detections, which presently is limited to two (\citealt{Weng2022}; see also \citealt{Jaron2024} for arguments that may question an origin of the pulses from \lsi{}). \citet{Massi2017} found that certain properties of the X-ray emission from \lsi{} are compatible with accreting black holes rather than neutron stars.

\subsection{Physical scenarios for periodic emission}\label{physsc}

The aim of the present work is to identify the physical periods in the radio emission from \lsi{} without the assumption of any specific physical scenario during or for the analysis. Nevertheless, we here briefly summarize the current status of the debate in the literature concerning possible physical processes behind the periodic signals from this source.

There is consensus about $P_1$ being the orbital period. It was first detected by \citet{Taylor1982} in the radio emission from \lsi{} (c.f.\ our discussion in Sect.~\ref{sec:TG82} here). A timing-analysis of the optical polarization presented by \citet{Kravtsov2020} seems to confirm that $P_1$ (rather than, e.g., $P_2$) is the orbital period. The debate starts over the question of which one of the two periods, $P_2$ (the quasi-orbital period) and ~$P_{\rm long}$ (the super-orbital period), is a physical period.

\citet{Massi2013} argued that $P_2$ is the intrinsic second period and that this period is  identical to the precession period of a relativistic jet. This interpretation is based on phase-referenced Very Long Baseline Array (VLBA) observations by \citet{Massi2012}, who showed that the core position of the radio brightness distribution of \lsi{} performs an ellipse on the sky that takes longer than one orbit ($\sim 27-28$~d) to return to its initial position. \citet{Wu2018} revisited the source with the VLBA and by analyzing the combined dataset, they determined a precession period that is in precise agreement with the estimates of $P_2$ from the different timing analysis studies mentioned above. These authors interpreted the observed variability of the radio brightness distribution as the precession of the radio core of a jet. The possibility of a precessing jet in \lsi{} had already been proposed, for instance, by \cite{mas04} (see also \citealt{gre89}). Coupled with  astrometric observations \citep{Massi2012, Wu2018}, it was  tested via the physical modeling of radio \citep{Massi2014} and GeV data \citep{Jaron2016}. In this scenario, the long-term period, $P_{\rm long}$, is the result of the interference between periodic mass-loading ($P_1$) and precession of a relativistic jet ($P_2$), giving rise to periodic changes in the Doppler boosting of the intrinsic emission, the beating between the orbital and precession periods resulting in the long-term modulation.

Several different explanations regarding the reality of $P_2$ and $P_{\rm long}$ have been put forth by \citet{Chernyakova2023}. These authors confirm the presence of $P_1$ and $P_2$ in the power spectrum of the GeV \gammaray{} emission from \lsi{}, as obtained by  means of aperture photometry of \fermilat{} observations\footnote{\citet{Chernyakova2023} decided to switch the order of the indices, placing $P_1$  as the quasi-orbital and $P_2$ as the orbital period. Throughout this article, we stick to the convention that $P_1$ is the orbital and $P_2$ is the quasi-orbital period, even when referring to their paper.}. Their new finding is that the emission at lower GeV energies (0.1-0.3~GeV) is modulated by $P_1$ only, while the higher energy emission (1-10~GeV) is modulated by $P_2$ and the long-term period, $P_{\rm long}$, without any sign of $P_1$ in the power spectrum (see their Fig.~2).
In their first interpretation of these results, they point out that the Keplerian velocity of the Be star decretion disk material at the truncation radius may well have a similar period (i.e., perhaps $P_2$) as the orbital period of the binary system itself, if the disk is truncated at a radius similar to the separation of the two components. In this case, $P_2$ may be an intrinsic period.
Their second scenario is that the Be star decretion disk is inclined with respect to the orbital plane, and that the pulsar's gravitational pull acts on the Be star disk, inducing precession. They point out that the precession period in this case is expected to be close to the orbital period. This means that  in this scenario, $P_2$, is also an intrinsic period.
In their third explanation, they attribute the long-term modulation to periodic growth and decay of the Be star disk. In this case, $P_{\rm long}$ would be an intrinsic period, while $P_2$ would then be interpreted as a beat period. 

A phenomenon observed in certain states of cataclysmic variables, involving two close periods and a longer one, is the super-hump oscillation \citep{Warner1995}. We here briefly mention the possible connection of this phenomenon with the presence of periodic signals  in \lsi{}. Super-humps are also observed in some neutron star and black hole low-mass X-ray binaries (\citealt{Kosenkov2018}, and references therein) and other similar systems with a white dwarf (e.g., Canis-Venaticorums, \citealt{Skillman1999}). In these cases, accretion disk deformations with super-orbital periods lead to quasi-orbital periods that are beats of the orbital and the super-orbital period. We note in passing that \lsi{} could be equivalent to a low-mass X-ray binary with the Be star's decretion disk playing the role of the accretion disk in these other sources and the compact object the role of the secondary low-mass star. The estimate from the mass ratio ($\sim 0.1$, see Fig.~8 in \citealt{Kosenkov2018}) matches the value of the quasi-orbital period ($P_2$) of \lsi{}.

Finally, there is an important possibility to account for, which is implicit for instance in the type of model presented by \cite{mar95} (see also references therein). In that work, the authors assumed that the Be disk properties change periodically with $P_{\rm long}$, inducing a consistent delay in orbital phase of the emission maximum with the same periodicity, as the interaction between the Be disk and the emitting region slightly changes orbit by orbit. In \cite{mar95}, this interaction was assumed to take place via accretion, but other types of interaction can easily fit in that framework \citep[see, e.g.,][and references therein, for the case of the compact object being a pulsar]{Dubus2013}. Thus, in this framework the emitting region physically changes in a complex but periodic manner and has two intrinsic periods in addition to the orbital one: $P_{\rm long}$ and $P_2$; the former associated to slow changes of the star or its disk, for example, and the latter due to the consequent small delay in the emission peak orbit by orbit. This kind of framework,  implying that $P_1$, $P_2$, and $P_{\rm long}$ are all physical and not the result of a beat, is agnostic with respect to the specifics of the compact object or even the different emitting regions involved (although, causally, the $P_2$ process would be a consequence of the $P_{\rm long}$ process).

\section{The data} \label{sec:data}

For the analysis presented in this paper, we made use of observational data from two long-term radio flux monitoring programs. The first program was carried out by the Green Bank Interferometer (GBI) during the second half of the 1990s \citep{Ray1997}. The second program has been carried out by the Owens Valley Radio Observatory (OVRO) since 2008 and is still on-going \citep{Richards2011}. In addition to that we revisited the data published by \citet{Taylor1982} from the beginning of the 1980s. These data are taken from their Table~1.

\begin{figure*}
  \includegraphics{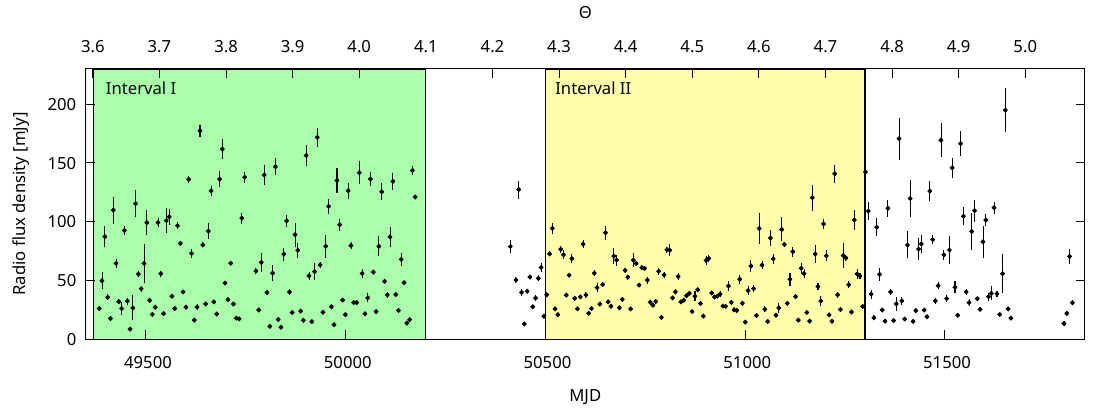}
  \caption{
    Radio light curve resulting from long-term monitoring of \lsi{} by the GBI at 8~GHz. The lower time axis is expressed in terms of MJD and the upper one in cycles of the long-term modulation, using $P_{\rm long} = 1667$~d and $T_0 = \mathrm{MJD}~43366.275$. These data have been averaged in time bins of width seven days. Two intervals are highlighted (green and yellow), which are analyzed individually, as explained in the text.
  }
  \label{fig:GBI8GHz-lc}
\end{figure*}

The GBI observed \lsi{} at~2 and~8~GHz simultaneously from January 27, 1994 until October 6, 2000. We select the observations at 8~GHz here, because the periodic signals that are subject of our investigation have been shown to become more significant toward higher radio frequencies \citep{Massi2016}. During this time span the source was observed almost daily with an average number of eight observations per day. This means that the resulting light curve has a sampling rate that is much higher than the Nyquist limit for the detection of the periodic signals that we are interested in. Furthermore, since \lsi{} is known to exhibit intraday micro-flaring activity, which can be characterized by quasiperiodic oscillations (\citealt{Jaron2017}; see also \citealt{Per97}), we decided to average the data in time-bins of width seven days. This sampling rate is still well above the Nyquist limit, but should be low enough to filter out any unwanted signals. The resulting time-series of averaged flux densities is plotted in Fig.~\ref{fig:GBI8GHz-lc}. In this plot, as in the other light curves presented in the following, the lower time axis is expressed in modified Julian day (MJD), and the upper axis in terms of long-term cycles $\Theta$ since the first radio detection of the source at $T_0 = 43366.275$\,MJD, using the value $P_{\rm long} = 1667$~d \citep{Gregory2002}. It is evident from this plot that there are two larger gaps in the data. The first gap occurs between MJD~50174 and MJD~50412 and the second gap toward the end of the data between MJD~51665 and MJD~51798. The two color-shaded intervals are used in the detailed analysis presented in Sect.~\ref{sec:results-int}. These intervals extend from MJD~49370 to MJD~50200 ($\Theta = 3.60-4.10$, interval~I in green) and from MJD~50500 to MJD~51300 ($\Theta = 4.28 - 4.76$, interval~II in yellow). These intervals cover 50 and 48\,\% of the length of the long-term modulation, respectively.

\begin{figure*}
  \includegraphics{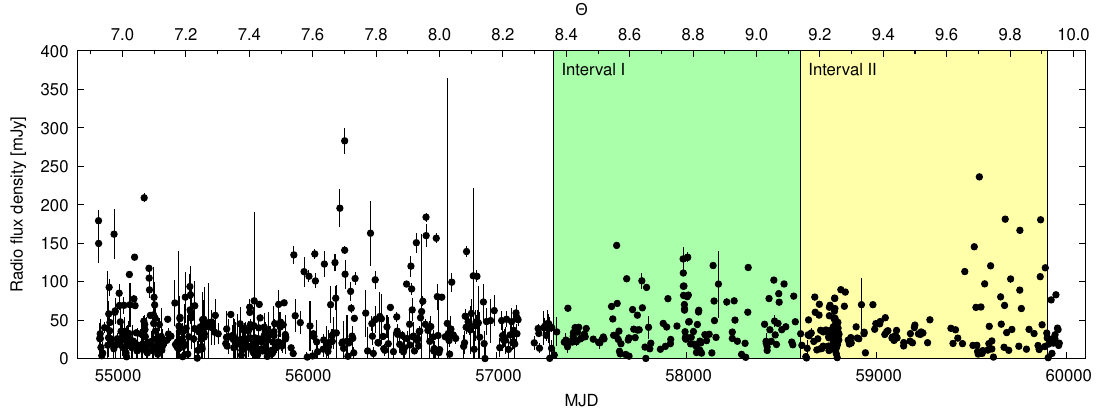}
  \caption{
    Radio light curve of \lsi{} observed by the OVRO at 15 GHz. The two color-shaded intervals are analyzed separately.
  }
  \label{fig:OVRO15GHz-lc}
\end{figure*}

The OVRO has been observing \lsi{} since March~18, 2009 at a radio frequency of 15~GHz, as part of a larger monitoring program of \textit{Fermi}-detected active galactic nuclei \citep{Richards2011}.
The light curve is shown in Fig.~\ref{fig:OVRO15GHz-lc}, where the upper time axis, expressed in terms of long-term cycles~$\Theta$, reveals that these are three full cycles of the long-term modulation. These are the same data that have also been subject of the analysis presented in \citet{Jaron2024}. Two intervals are highlighted in green and yellow, respectively. Interval~I extends from MJD~57300 to MJD~58600, which is $\Theta = 8.36-9.14$, namely, 78\,\% of the long-term period. Interval~II extends from MJD~58600 until MJD~59900, which is $\Theta = 9.14-9.92$, namely, 78\,\% of $P_{\rm long}$. The individual analysis of these two intervals is presented in Sect.~\ref{sec:results-int}.

\section{Methods} \label{sec:methods}

\subsection{Lomb-Scargle periodogram}

The Lomb-Scargle periodogram \citep{Lomb1976, Scargle1982} is a powerful and
well-established method to detect periodic signals in unevenly sampled
astronomical time-series \citep{VanderPlas2018}. \citet{Zechmeister2009}
presented a potentially more robust modification, which they refer to as
generalized Lomb-Scargle (GLS) periodogram. For the timing analysis presented
in this paper, we used the Python implementation of the GLS that is part of
the PyAstronomy package \citep{Czesla2019}.

\subsection{Period determination}

We determine the values of periodic signals found in the data by means of GLS by fitting the periodogram with an appropriate function. We empirically used the superposition of two Gaussians
\begin{equation} \label{eq:Gauss}
  f(\nu) = a_1\mathrm{exp}\left(-\frac{1}{2}\frac{(\nu - \nu_1)^2}{\sigma_1^2}\right) + a_2\mathrm{exp}\left(-\frac{1}{2}\frac{(\nu - \nu_2)^2}{\sigma_2^2}\right),
\end{equation}
with amplitudes $a_{1,2}$, center frequencies $\nu_{1,2}$, and standard deviations $\sigma_{1,2}$. As outlined in \citet{VanderPlas2018}, we plot and process the periodograms in terms of frequency, rather than period. We report period values as $P = 1/\nu \pm \sigma/\nu^2,$ resulting from Gaussian error propagation and take the parameter~$\sigma$ from the fit of Eq.~(\ref{eq:Gauss}) as the uncertainty of the frequency,~$\nu$. In the case of a one-peaked profile we omit the second Gaussian in Eq.~(\ref{eq:Gauss}).

\section{Results} \label{sec:results}

\subsection{Timing-analysis of restricted intervals} \label{sec:results-int}

\begin{figure*}
  \includegraphics{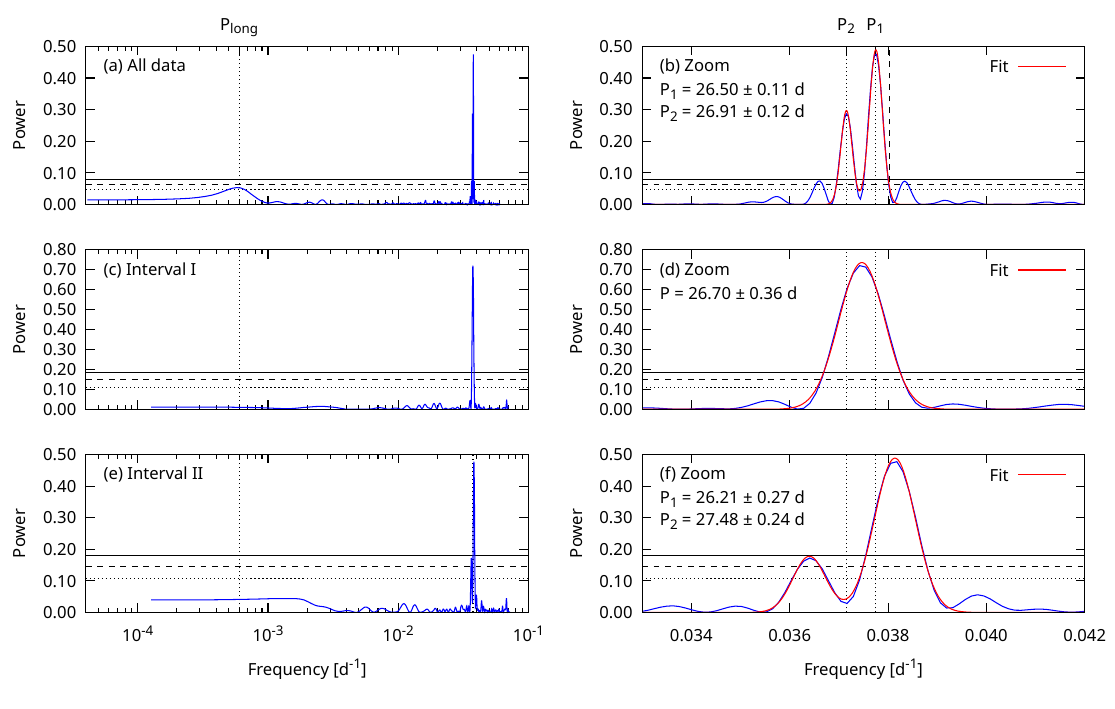}
  \caption{
    GLS periodograms of the GBI 8~GHz data. The horizontal lines indicate FAP levels of 10\,\% (dotted), 1\,\% (dashed), and 0.1\,\% (solid).
    (a)~Periodogram of the full light curve shown in Fig.~\ref{fig:GBI8GHz-lc}. A peak at $P_{\rm long}$ is present along with a feature at $P_1$.
    (b)~Zoom onto the region around the orbital period. A distinct two-peaked profile is present. The vertical dashed lines marks the position of the possible third period, $P_3$. The result of fitting the periodogram with Gaussian functions is shown as the red solid curve. We report the uncertainties of the resulting periods using the $1\sigma$ standard deviations of these Gaussians.
    (c)~Periodogram of interval~I (green area in Fig.~\ref{fig:GBI8GHz-lc}). The peak at $P_{\rm long}$ is absent.
    (d)~The zoom onto the orbital region reveals the presence of only one peak, at $P_{\rm average}$.
    (e)~Periodogram of interval~II (yellow area in Fig.~\ref{fig:GBI8GHz-lc}).
    (f)~Zoom onto the orbital region. There is a two-peaked profile, with the periods being largely compatible with the literature values of $P_1$ and $P_2$. 
  }
  \label{fig:GBI8GHz-ls}
\end{figure*}

In Fig.~\ref{fig:GBI8GHz-ls}, the GLS periodograms of the different time intervals of the GBI 8~GHz data are plotted. The time intervals are those highlighted in the light curve of Fig.~\ref{fig:GBI8GHz-lc}. The periodogram shown in panel~(a) of Fig.~\ref{fig:GBI8GHz-ls} was calculated using the entire GBI 8~GHz data. The power spectrum itself is very clean. The most powerful feature is at the position of the orbital period. Besides that, there is also a minor peak at the position of the long-term period, indicated by the vertical dotted line. Zooming in on the feature at the orbital period in panel~(b) reveals that there are two peaks, precisely at the positions of the well-known periods $P_1 = 26.4960$~d \citep{Gregory2002} and $P_2 = 26.926$~d \citep{Wu2018}, as indicated by the  vertical dotted lines in the plot. Levels of false alarm probability (FAP) are shown as dotted (10\,\%), dashed (1\,\%), and solid (0.1\,\%) horizontal lines in this and all the other periodograms shown throughout this article. Both the peak at $P_1$ and the one at $P_2$ are well above the solid line, representing $\mathrm{FAP} \ll 0.1$\,\%. The side-lobes are below that line. The position of the possible third period,~$P_3 = 26.3$~d \citep[as reported by][]{Zhang2024},  is indicated by the vertical dashed line in this plot .  It is, however, evident that the side-lobe on the right-hand side of the peak at $P_1$ is not at the position of $P_3$. A fit of a two-peaked Gaussian function to the periodogram results in $P_1 = 26.50 \pm 0.11$~d and $P_2 = 26.91 \pm 0.12$~d, in perfect agreement with the values reported in the literature, in particular \citet{Massi2013}, who performed timing analysis of the same dataset. We repeated their analysis here, employing the modern GLS algorithm, which was not available during that time.

In the following, we inspect the periodograms of the two intervals~I and~II, which are significantly shorter than one cycle of the super-orbital modulation period~$P_{\rm long} = 4.6$~years. These two intervals are highlighted by the color-shaded areas in the GBI~8~GHz light curve shown in Fig.~\ref{fig:GBI8GHz-lc}. Interval~I is highlighted in green and extends from MJD~49370 to MJD~50200, which are 830~days or 50\,\% of the super-orbital period. This interval is taken from the maximum of the long-term modulation pattern. Interval~II extends from MJD~50500 until MJD~51300, which is 800~days or 48\,\% of the long-term modulation. This interval was chosen to include the minimum of the long-term modulation and extends symmetrically around it.

The periodogram for interval~I is shown in panel~(c) of Fig.~\ref{fig:GBI8GHz-ls} and is basically featureless besides a strong peak at the position of approximately the orbital period. A zoom onto this region, shown in panel~(d), reveals that there is only one peak. This peak is well above the solid line, indicating $\mathrm{FAP} \ll 0.1$\,\%. The very small side-lobes are well below the dotted line, meaning that they can be considered insignificant. The single peak has a width that includes both the nominal values of $P_1$ and $P_2$. The Gaussian fit results in $P = 26.70 \pm 0.36$~d, which is exactly the average of $P_1$ and $P_2$. This finding is in agreement with \citet{Ray1997}, who analyzed the interval MJD~49379-50136 (very similar to our interval~I), and determined $P = 26.69 \pm 0.02$~days. These authors, however, applied a cross-correlation method (that was also used by \citealt{Taylor1982}), which might explain their smaller uncertainty. We used the standard deviation~$\sigma$ of the fitted Gaussian as the frequency uncertainty, which we consider a conservative choice.

Figure~\ref{fig:GBI8GHz-ls}~(e) shows the GLS periodogram for interval~II, namely, the yellow-shaded area in Fig.~\ref{fig:GBI8GHz-lc}. While there is some power at the position of the long-term period, this cannot be regarded as a distinct peak, compared to the plot in panel~(a). Zooming in on the feature at the orbital period, panel~(f) reveals a two-peaked profile. This profile is very well fitted by a two-peaked Gaussian function. The resulting period values $P_1 = 26.21 \pm 0.27$~d and $P_2 = 27.48 \pm 0.24$~d are largely compatible with the nominal values of these periods, given their large uncertainties. The peak at the higher frequency is well above the solid line, while the lower frequency peak touches this line. Besides these two peaks, there are not any other significant features in this frequency interval. We point out that this is the detection of a two-peaked profile in a time interval that is significantly shorter than the long-term period itself. The fact that the peaks are broader, compared to the ones shown in panel~(b), is explained by the fact that the frequency resolution of the GLS method is inversely proportional to the length of the time series that is being analyzed.

\begin{figure*}
  \includegraphics{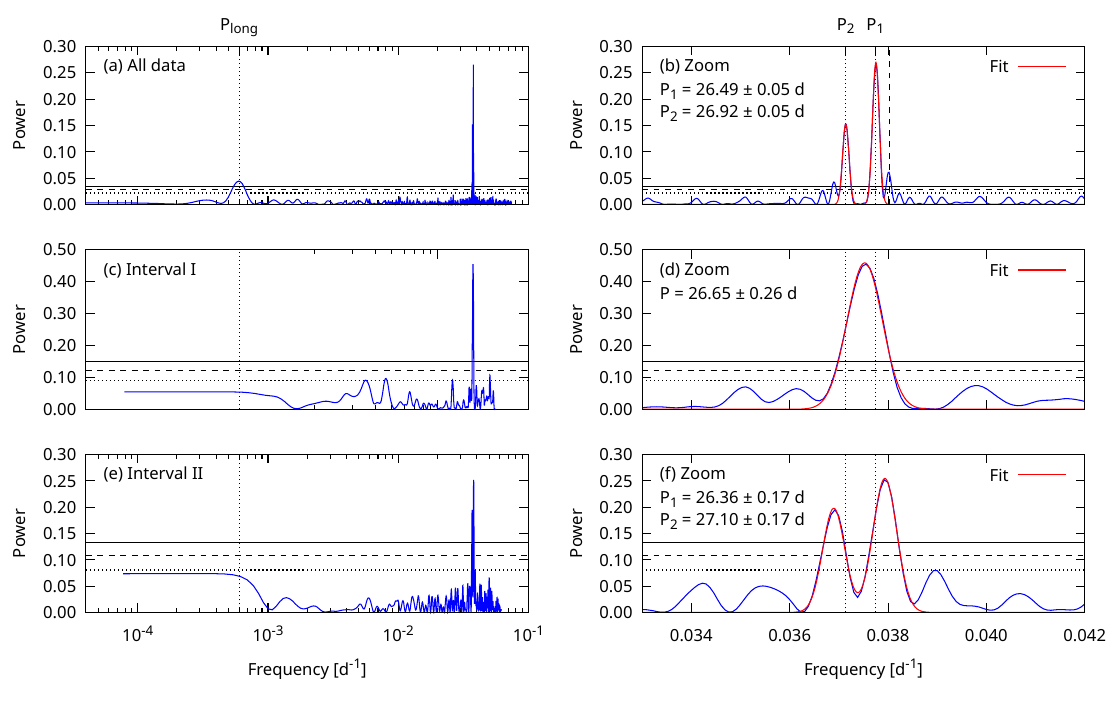}
  \caption{
    GLS periodograms of the OVRO 15 GHz data.
    (a)~Periodogram of the entire light curve shown in Fig.~\ref{fig:OVRO15GHz-lc} \citep[c.f.][]{Jaron2024}.
    (b)~Zoom onto the region around the orbital period. A distinct two-peaked profile is present. The vertical dashed line marks the position of $P_3$, as reported by \citet{Zhang2024}. A significant feature is indeed detected at this position.
    (c)~Periodogram of interval~I (green-shaded area in Fig.~\ref{fig:OVRO15GHz-lc}). The only significant feature is at approximately the orbital period.
    (d)~Zoom. There is only one peak.
    (e)~Periodogram of interval~II (yellow-shaded area in Fig.~\ref{fig:OVRO15GHz-lc}). The only significant feature is at approximately the orbital period.
    (f)~Zooming in on the orbital range reveals a two-peaked profile.
  }
  \label{fig:OVRO15GHz-ls}  
\end{figure*}

In the following we apply an analogous analysis to the radio data taken by the OVRO at 15~GHz. The results of the GLS analysis are presented in Fig.~\ref{fig:OVRO15GHz-ls}. The periodogram of the full dataset, subject of panel~(a), reveals a very clean spectrum with the only significant feature being a peak at $P_{\rm long}$ and another one at the orbital period. The zoom-in panel~(b) shows a very distinct two-peaked profile, that is very well fitted by a two-peaked Gaussian and results in $P_1 = 26.49 \pm 0.05$~d and $P_2 = 26.92 \pm 0.05$~d, in full agreement with the well-established values of these two periods. This is a repetition and confirmation of the analysis and results presented in \citet{Jaron2024}. In addition to their analysis, however, we here also recognize that both side-lobes in Fig.~\ref{fig:OVRO15GHz-ls}\,(b), namely, the one on the left-hand side of~$P_2$ and the one on the right-hand side of~$P_1$, are above the solid line that indicates that these features have FAP$\ll 0.1$\,\%. In particular, the feature on the right-hand side of~$P_1$ is exactly at the position of $P_3$, reported by \citet{Zhang2024}. This is a confirmation of the presence of a significant peak in the periodogram of radio data from \lsi{} at the position of $P_3$. The non-detection of such a feature in the GBI data is most likely the consequence of the shorter time span covered by the GBI data compared to the OVRO data. The longer time span of the OVRO light curve results in an increased frequency resolution, enabling the detection of $P_3$. An investigation of the nature of this feature is the subject of the analysis of synthetic data in Sect.~\ref{sec:results-synth}.

In the periodogram of interval~I (green-shaded area in Fig.~\ref{fig:OVRO15GHz-lc}) shown in panel~(c), the peak at $P_{\rm long}$ is not so well pronounced anymore. The only significant peak is at approximately the orbital period. The zoom-in panel~(d) shows only one peak, which is located in the middle between $P_1$ and $P_2$. This peak is fitted by a single Gaussian and has a center period of $P = 26.65 \pm 0.26$~d. This result is very similar to the one that was obtained from analyzing interval~I of the GBI data. The intervals~I of GBI and OVRO share the property of not including a long-term minimum.

Fig.~\ref{fig:OVRO15GHz-ls}~(e) shows the periodogram for interval~II (yellow-shaded area in Fig.~\ref{fig:OVRO15GHz-lc}). Again, the only significant feature is located at the position of the orbital period. The zoom-in panel~(f) reveals a two-peaked profile, very well fitted by two Gaussians, which result in $P_1 = 26.36 \pm 0.17$~d and $P_2 = 27.10 \pm 0.17$~d, both in agreement with the values reported in the literature. Also in this case, the result is similar to the one obtained for interval~II of the GBI data, the difference being that $P_2$ is detected with higher significance in the OVRO data. Both the intervals~II of GBI and OVRO share the property of including a minimum of the long-term modulation.

In this section we have demonstrated that it is possible to detect the two-peaked profile, consisting of $P_1$ and $P_2$, in observational datasets that are significantly shorter than one cycle of the long-term modulation. This is, however, only possible when choosing an interval that contains a minimum of the long-term modulation. This is the case for the intervals labeled ``II.'' Otherwise, only one peak is detected that is located between~$P_1$ and~$P_2$ and is best in agreement with the average period~$P_{\rm average} = 26.7$~d.

\subsection{Interlude: Reasoning behind the\ \citet{Taylor1982} detection of only $P_1$} \label{sec:TG82}

The first detection of a periodic signal in the radio emission from \lsi{} was reported by \citet{Taylor1982}. By performing a cross-correlation analysis on their radio data, resulting from observations at~5 and 10~GHz, these authors determined a period value of $26.52 \pm 0.04$~d. This value is significantly different from $26.69 \pm 0.02$~d, which was obtained by \citet{Ray1997}, who analyzed a different dataset of radio observations from the same source and at similar radio frequencies (2 and~8~GHz), but observed several years later (i..e, the beginning of the GBI light curve, shown in Fig.~\ref{fig:GBI8GHz-lc} here). In the following we revisit the issue of why these authors came to these competing results. In particular, we address the question why \citet{Taylor1982} only detected a period that is very well in agreement with the orbital period $P_1$. Based on the results presented in the previous subsection, one might expect that they should have detected either a period that is compatible with $P_{\rm average}$, as \citet{Ray1997} did, or that they could also have detected the period~$P_2$ along with, or possibly instead of, the orbital period~$P_1$. Their result should have depended on the coverage of long-term modulation phases (the long-term modulation itself had not been detected before \citealt{ParedesPhD} and \citealt{gre89}).

\begin{figure}
  \includegraphics{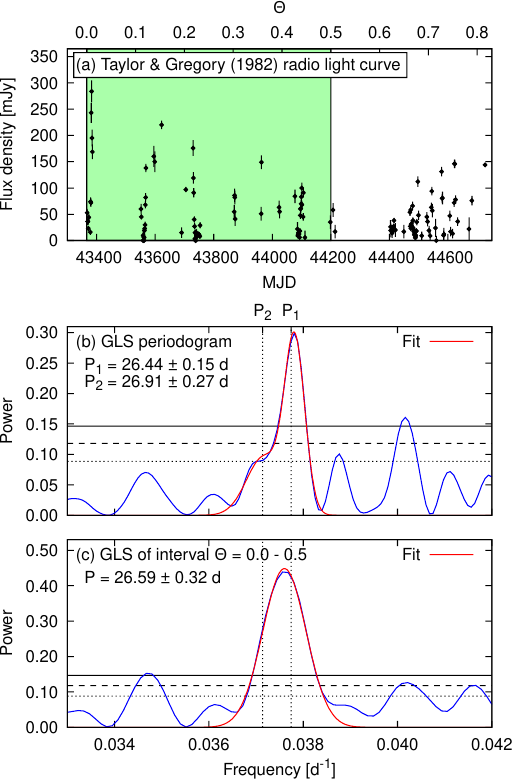}
  \caption{
    Radio observations of \lsi{} taken from \citet{Taylor1982}.
    (a)~Light curve at 5 and 10.5~GHz. An interval for separate analysis is shaded in green.
    (b)~Periodogram of the full light curve. The most powerful peak is consistent with the orbital period. The fit to the periodogram, however, requires the superposition of two Gaussians.
    (c)~Periodogram of super-orbital interval $\Theta = 0.0 - 0.5$ (green-shaded area in panel~a). The one-peaked profile is shifted toward the middle of the positions of $P_1$ and $P_2$ and is fitted by a single Gaussian.
  }
  \label{fig:Taylor1982}
\end{figure}

In Fig.~\ref{fig:Taylor1982}~(a), the flux measurements listed in Table~1 of \citet{Taylor1982} are plotted. The lower time axis is given in MJD while the upper one is given in units of the long-term modulation phase,~$\Theta$. The long-term flux modulation is discernible by eye, coming from a maximum at $\Theta = 0.0$, having a minimum at around $\Theta = 0.5-0.6$ (where there is unfortunately a gap in the data), after which the fluxes begin to increase again. This means that we are dealing here with an interval of radio observations of \lsi{} that is shorter than one full cycle of the long-term modulation but at the same time includes a long-term minimum. The green-shaded area is an interval that was later analyzed individually, as described later in this work.

Panel~(b) of Fig.~\ref{fig:Taylor1982} shows the GLS periodogram of the entire light curve shown in panel~(a), zoomed into the region around the orbital period. The strongest feature is indeed a peak that is in agreement with the orbital period~$P_1$, as indicated by the right vertical dotted line. The position of an expected peak at $P_2$ is marked by the left dotted line, but there is not any distinct peak present at that position. When fitting the periodogram of that region, we however need the superposition of two Gaussian functions in order to fit the extension of the peak to the left of it. The resulting value for the orbital period is $P_1 = 26.44 \pm 0.15$~d, in agreement with the value originally reported by \citet{Taylor1982}. The value for the other period is $P_2 = 26.91 \pm 0.27$~d. We do not claim that this is a significant detection of $P_2$ in this dataset, but we point out that it is necessary to include a second Gaussian in the fitting, which might indicate that with this dataset we are on the edge for the detection of this period.

Next, we restricted the analysis to the interval marked by the green-shaded area in panel~(a). The limits of this interval were chosen such that it spans exactly the super-orbital phase range $\Theta = 0.0 - 0.5$. As can be seen in the figure, it does not include the minimum of the long-term modulation. The GLS periodogram of these data is shown in panel~(c). There is only one peak, which is shifted toward the middle of the positions of $P_1$ and $P_2$, compared to the peak in panel~(b). Fitting this peak does not require a second Gaussian and results in $P = 26.59 \pm 0.32$~d. Within its uncertainties, this value is in agreement with both $P_1$ and $P_2$ and in particular with the period that was found by \citet{Ray1997}, namely, with~$P_{\rm average}$.

We tentatively come to the following conclusion about why \citet{Taylor1982}  only ended up detecting $P_1$, and not $P_2$ or $P_{\rm average}$. The time interval of data which they analyzed covered a super-orbital phase range of $\Theta = 0.0 - 0.8$, which is less than one full cycle of the long-term modulation. Since this interval includes the minimum of the long-term modulation, always occurring at $\Theta \approx 0.5$ (see \citealt{Jaron2024} and references therein), it should in principle enable the distinction between $P_1$ and $P_2$ by the timing analysis. This is the reason why the GLS gives a well-pronounced peak at $P_1$ in the periodogram shown in Fig.~\ref{fig:Taylor1982}~(b). The necessity of fitting the superposition of two Gaussians to this peak, which further results in period values in agreement with the nominal values of $P_1$ and $P_2$ can be interpreted as at least a hint that $P_2$ is also present in these data, however at a level that is below the detection limit. Since in almost all periodograms available for this source and this radio frequency range, the period $P_1$ usually gains a higher power than $P_2$, it is the most likely outcome that only $P_1$ is detected under these circumstances. The time series analyzed by \citet{Taylor1982} is long enough to resolve $P_1$ (and $P_2$), but it is not long enough, or does not have beneficial sampling properties, to give enough power to $P_2$. Another possibility, that cannot be ruled out based on these results, is that the power of $P_2$ is intrinsically lower during the time span of these observations than it is in the light curves observed at later times (GBI, OVRO). This would imply a physical origin, intrinsic to the processes at work in \lsi{}, for the non-detection of $P_2$ in these data. In any case, $P_1$ is certainly the dominating period here and an identification with the orbital period of the binary system seems natural.

\subsection{Filtering out $P_1$ and $P_2$} \label{sec:filter}

\begin{figure}
  \includegraphics{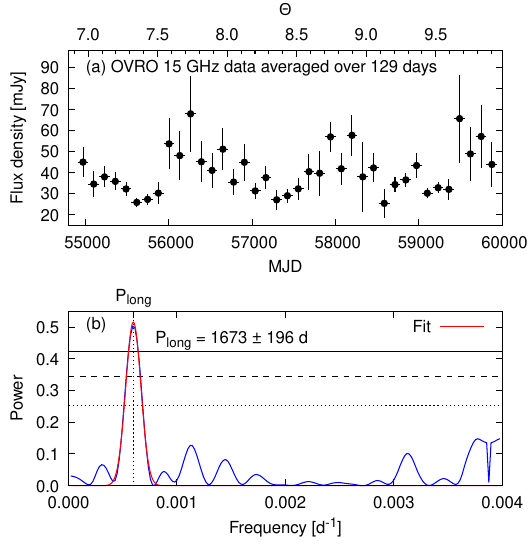}
  \caption{
    (a) OVRO 15 GHz data (see Fig.~\ref{fig:OVRO15GHz-lc}) averaged in time-bins of width 129~d.
    (b) GLS periodogram of the averaged data. There is only one significant peak, at a period of $1673 \pm 196$~d, in full agreement with the long-term period of the source.
  }
  \label{fig:OVRO-avg}
\end{figure}

In the following, we investigate if the long-term period,~$P_{\rm long}$, is still detected after filtering out the short-term periods $P_1$ and $P_2$ from the data. For this purpose we make use of the OVRO data (plotted in Fig.~\ref{fig:OVRO15GHz-lc}), because these data span three cycles (i.e., more than GBI) of the long-term modulation. We chose to average the data in equidistant time bins of width 129~d, because this is a value that is below the Nyquist limit for the detection of $P_1$ and $P_2$ but still well above the one for the detection of $P_{\rm long}$. In addition to that, this value is not related to any of these three periods by a simple mathematical multiplication or division. The averaged OVRO data are plotted in Fig.~\ref{fig:OVRO-avg}~(a). It is visible by eye that the averaged fluxes are still subject to a long-term modulation.

To test whether the long-term period was still included in these data, we applied the GLS method to the averaged dataset. The result of this is shown in Fig.~\ref{fig:OVRO-avg}~(b), where the full range of frequencies left over after the averaging is plotted. The only significant peak is, indeed, at the position of the nominal value of the long-term period, as indicated by the vertical dotted line. The fact that this peak exceeds the horizontal solid line means that its FAP is below 0.1\,\%. The fit of this peak with a single Gaussian function gives $P_{\rm long} = 1673 \pm 196$~d, a value that is in very good agreement with the other values of the long-term period reported in the literature. This means that the periodic long-term modulation survives after filtering out the two short-term periods by the means of averaging below their Nyquist limit.

\subsection{Synthetic data} \label{sec:results-synth}

\begin{figure*}
  \includegraphics{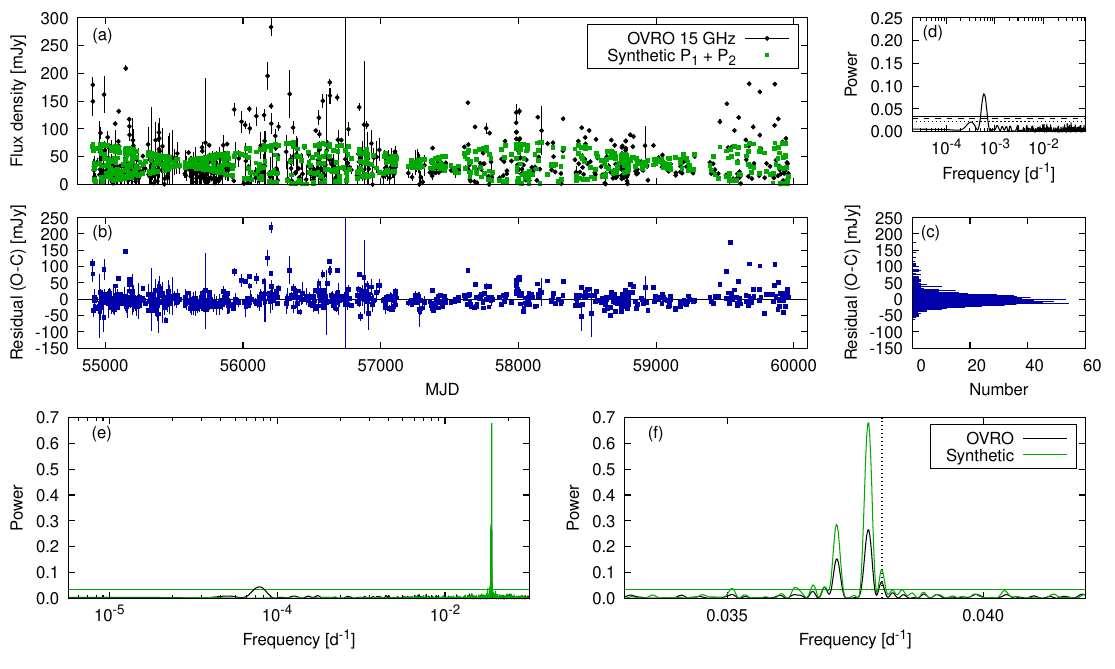}
  \caption{
    Synthetic data and their analysis.
    (a)~The original OVRO observations are plotted as black filled circles. The synthetic data, containing the periods $P_1$ and $P_2$, are plotted as green filled squares.
    (b)~Residuals (observed minus computed).
    (c)~Histogram of the residuals.
    (d)~GLS periodogram of the residuals, showing a significant peak at $P_{\rm long}$.
    (e)~Full GLS periodogram of the observational (black) and the synthetic data (green). The peak at $P_{\rm long}$, present for the observational data, is missing for the synthetic data. The horizontal solid green line indicates the 0.1\,\% FAP level of the synthetic data.
    (f)~Zoom onto the orbital frequency range. The two periods, $P_1$ and $P_2$, are present for both the observational and the synthetic data. The nominal position of $P_3 = 26.3$~d is marked by the vertical dotted line. A feature is present at that position for both the observational and the synthetic data.
  }
  \label{fig:synthP1P2}
\end{figure*}

\begin{figure*}
  \includegraphics{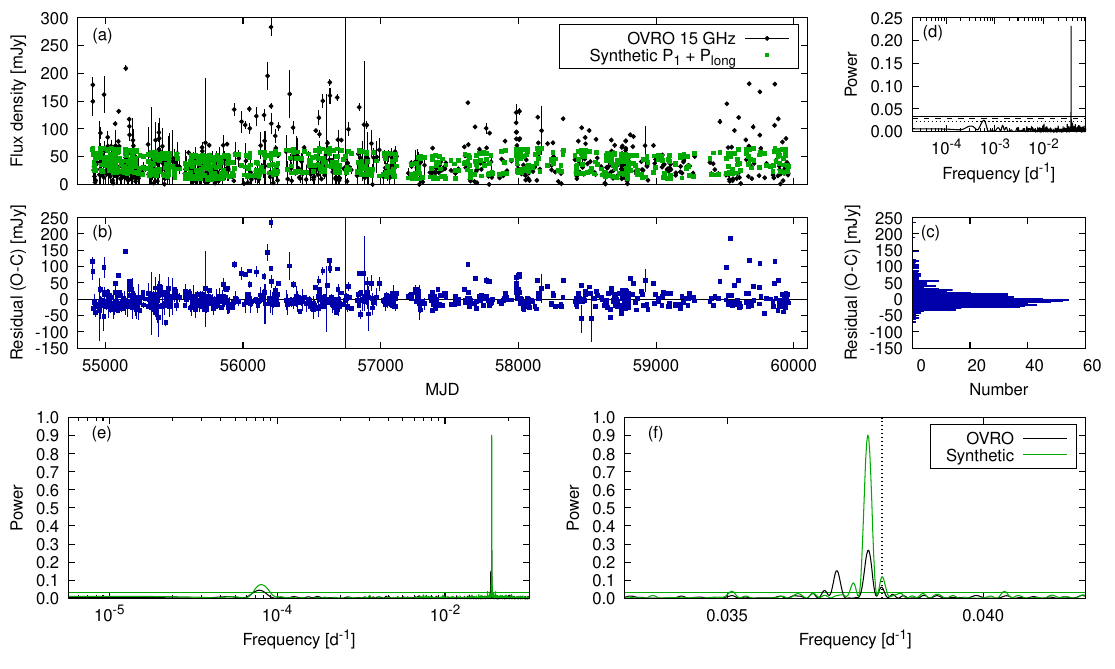}
  \caption{
    Synthetic data and their analysis.
    (a) Synthetic data contain the periods $P_1$ and $P_{\rm long}$.
    (b)~Residuals (observed minus computed).
    (c)~Histogram of the residuals.
    (d)~GLS periodogram of the residuals, showing a significant peak at $P_2$.
    (e)~Full GLS periodogram, the peak at $P_{\rm long}$ is present for both the observational and the synthetic data.
    (f)~Zoom onto the orbital frequency range. The peak at $P_1$ is reproduced by the synthetic data, but besides that, the periodograms differ substantially between observational and synthetic data.
  }
  \label{fig:synthP1Plong}
\end{figure*}

\begin{figure*}
  \includegraphics{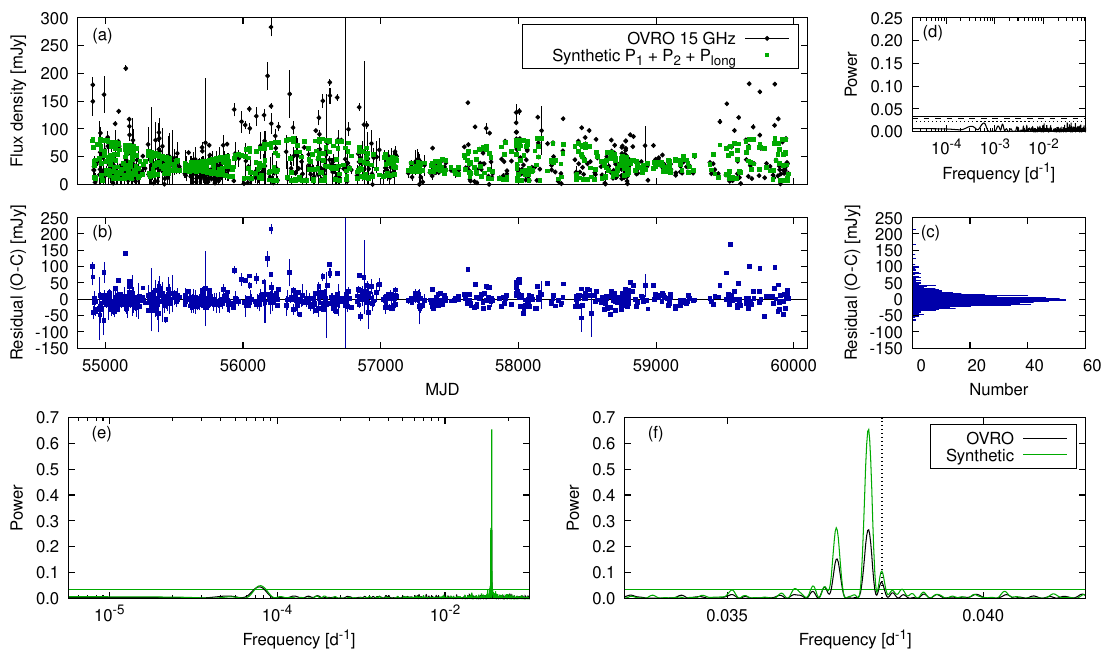}
  \caption{
    Synthetic data and their analysis.
    (a)~ Synthetic data contain all three periods, $P_1$, $P_2$, $P_{\rm long}$.
    (b)~Residuals (observed minus computed).
    (c)~Histogram of the residuals.
    (d)~GLS periodogram of the residuals, showing a featureless spectrum, demonstrating that the residuals contain only noise.
    (e)~In the full periodogram, the peak at $P_{\rm long}$, present in the observational data, is reproduced by the synthetic data.
    (f)~Zoom onto the orbital frequency range. The peaks at the two periods, $P_1$ and $P_2$, are present for both observational and synthetic data. For both datasets, there is also a significant feature at the position of $P_3$, as indicated by the vertical dotted line.
  }
  \label{fig:synthP1P2Plong}
\end{figure*}

Here, we address the question how the characteristic features of the periodogram of the radio emission from \lsi{} can be understood as the result of the interference of a few  intrinsic sinusoidal periodic signals. For this purpose, we approximated the OVRO radio light curve (Fig.~\ref{fig:OVRO15GHz-lc}) by a most simple mathematical function. By following this approach, we do not impose any assumption about the physical processes at work in the astrophysical system. We refer to this kind of generated data as synthetic data. 

We consider the function,
\begin{equation} \label{eq:synth}
  f(t) = \sum_{i = 1}^{N} a_i\sin\left(\frac{2\pi}{P_i}\cdot t - \phi_i\right) + c,
\end{equation}
which is the superposition of~$N$ sine functions with amplitudes,~$a_i$, periods,~$P_i$, and phase-origins,~$\phi_i$. In addition we include an overall constant offset~$c$. In our analysis, $N$ is either 2 or 3, and the a priori values for the periods,~$P_i$, are the nominal values of $P_1 = 26.5$~d, $P_2 = 26.9$~d, and $P_{\rm long} = 1667$~d, respectively. We used the SciPy Python package \citep{Virtanen2020} for a least-squares adjustment to the data. In all cases presented below, we left all parameters free for the fit. We did not impose any additional constraints on the fit, in particular, we did not require $f(t)$ to be strictly positive, nor did we clip any negative values of it, by either omitting these negative values or setting them to zero. Clipping the data would bias the data and introduce unwanted frequencies into the spectrum \citep{VanVleck1966}, which we avoid with our approach. 

The approach that we follow here bears a certain similarity to the analysis presented in \citet{Massi2013} but is significantly different in two aspects. The first aspect is that the functions used by these authors (their Eqs.~1-3), with the parameters used for their analysis, also return negative values. These authors did not include a constant offset in their functions (as the parameter~$c$ in our Eq.~\ref{eq:synth}), but instead put negative function values to zero (as apparent from their Fig.~2, panels~c, e, and~g, and confirmed by F.~Jaron, pers.~comm.). We follow a different approach here by including a constant offset and by not putting any function values to zero. The second aspect is that \citet{Massi2013}, in their analysis of the effect of the long-term period, used the model of amplitude modulation of the orbital period by the long-term period. This choice was made in response to the state of the discussion in the literature in the time of writing their paper. We here restrict ourselves strictly to the superposition of sine functions of the form of Eq.~(\ref{eq:synth}), and do not further modify the function values. These differences in the analyses, which are not directly comparable with each other, explain any differences in the results that we obtain here compared to those obtained by \citet{Massi2013}.

We first assumed a function $f(t)$ consisting of two sine waves at periods $P_1 = 26.5$~d and $P_2 = 26.9$~d. Fitting this function resulted in the parameters listed in Table~\ref{tab:fitresults} (Model: $P_1$ + $P_2$). The synthetic data, resulting from the evaluation of $f(t)$ at the observational epochs of the OVRO data, are plotted as green filled squares in Fig.~\ref{fig:synthP1P2}~(a). For comparison, the original OVRO data are plotted in black. The reduced $\chi^2$ has a value of 49. This is by no means a good fit to the data, but the closest that one can get with this simplistic approach. Despite this simplicity, it is worth of note how accurate the synthetic data follow the long-term modulation of the observational data, and how precise especially the shapes of the minima of the long-term modulation are reproduced. The residuals, in terms of observed minus computed values, are plotted in panel~(b). The histogram of residuals in panel~(c) shows a distribution centered at zero, albeit with a slight bias toward the positive direction. The weighted root mean square (WRMS) of these residuals is 20.7\,mJy.  Panel~(d) shows the GLS periodogram of the residuals. The only significant feature in this periodogram is a peak at the position of $P_{\rm long}$, everything else is below the dotted line (i.e., FAP > 10\,\%) and can be considered as noise. The GLS periodogram of the synthetic data for the full frequency range is plotted in panel~(e). The horizontal green solid line indicates FAP$= 0.1$\,\% (also in panel~f). The peak at the orbital period is present for both observational and synthetic data, but the peak at $P_{\rm long}$ is only present for the observational data.  It might be surprising that the beat period ($P_{\rm long}$) does not appear as a peak in the periodogram of these synthetic data. We have tested this behavior with other datasets, and the GLS method indeed only results in a periodogram that features peaks at $P_1$ and $P_2$, as long as these periods relate to pure sine waves, that are symmetric with respect to an average value. Breaking this symmetry can introduce the beat period in the periodogram, but we explicitly avoid this here. A justification for our sinusoidal approach comes from the fact that the residuals are indeed noise-limited (see panels b and c). Zooming in on the range around the orbital period in panel~(f), the two-peaked profile, consisting of $P_1$ and $P_2$, is reproduced by the synthetic data. The heights of the peaks are larger for the synthetic data, which is well explained by the absence of noise in these data. The nominal position of the $P_3 = 26.3$~d period, claimed by \citet{Zhang2024}, is marked by the vertical dotted line. A small peak is indeed present at this position for both observational and synthetic data. The peak is even stronger for the synthetic data than for the observational ones, both exceeding a FAP level of 0.1\,\%. This is an indication that $P_3$ can be explained as a side-lobe resulting from the beating between $P_1$ and $P_2$, and does not require the physical presence of the period~$P_3$ itself.

The second function that we fit contained the periods $P_1$ and $P_{\rm long}$. The fitted parameters are listed in Table~\ref{tab:fitresults} (Model: $P_1$ + $P_{\rm long}$). The resulting synthetic data are plotted in panel~(a) of Fig.~\ref{fig:synthP1Plong}. It is immediately visible that, while the synthetic data do follow the long-term trend, the detailed shape of the light curve is not so well reproduced as was the case for the previous example. The reduced $\chi^2$ has a value of 58.8, which is larger than in the previous case, but not by as much as one might expect from the qualitative difference of the results. The WRMS of the residuals, shown in panels~(b) and~(c), is 22.7\,mJy. The periodogram of the residuals in panel~(d) has a significant peak only at the position of $P_2$. The full GLS periodogram of the synthetic data in panel~(e) shows a peak at the long-term period. However, the zoom-in panel~(f) reveals that besides the peak at $P_1$, the shape of the overall profile is not reproduced at all by the synthetic data. In particular, the peak at $P_2$ is completely missing. There is, however, also in this case a significant peak at $P_3$. This shows that such a peak can also be explained by the beating between $P_1$ and $P_{\rm long}$, meaning that also in a scenario consisting of these two periods only, a physical $P_3$ is not required. The beating of $P_1$ with $P_{\rm long}$ results in a second side-lobe of $P_1$ that is smaller and located on the left-hand side of the peak. Such a feature is, however, completely absent in the periodogram of the observational data. Based on this result, we rule out the possibility that $P_2$ is a secondary period resulting from the beating between $P_1$ and $P_{\rm long}$.

Lastly, we fit a function that consisted of three sine waves at all three periods, namely, $P_1$, $P_2$, and $P_{\rm long}$, the results of which are shown in Fig.~\ref{fig:synthP1P2Plong}, and the fitted parameters can be found in Table~\ref{tab:fitresults} (Model: $P_1$ + $P_2$ + $P_{\rm long}$). The similarity between observational and synthetic data, shown in panel~(a), is improved over the first example, which contains only $P_1$ and $P_2$. The reduced $\chi^2$ in this case is 46.4, which is indeed the lowest value of all three examples. We do not expect a $\chi^2$ of unity for this simplistic mathematical approach and the amount of stochastic fluctuations coming from the source itself. The residuals in panels~(b) and~(c) are slightly improved as well, with a WRMS of 20.1\,mJy. The periodogram of the residuals in panel~(d) is basically featureless, remaining below the dotted line, which indicates $\mathrm{FAP} > 10$\,\%. This means that the residuals indeed only contain noise, most of which is coming from the source itself, explaining the relatively large values of $\chi^2$. In the full periodogram, subject of panel~(e), the peak at $P_{\rm long}$ is well reproduced and the zoomed-in panel~(f) shows that the two-peaked profile is there, along with a feature at $P_3$. In fact, the entire shape of the observational periodogram is remarkably well reproduced by the synthetic data. 

In summary, in order to reproduce all aspects of the periodogram with a simplistic approach as presented here, it is necessary to include all three periods (i.e., $P_1$, $P_2$, and $P_{\rm long}$) in the fitted function. This leads us to tentatively conclude that all of these three periods are true intrinsic periods of the binary system \lsi{}. The period $P_3$ is easily reproduced in all three cases as a simple side-lobe resulting from the interference of the intrinsic periods and we are inclined to rule out a physical reality of this period. However, we confirm that this is indeed a significant feature of the periodogram of radio data from \lsi{}, in agreement with the findings of \citet{Zhang2024}.

\begin{figure}
  \includegraphics{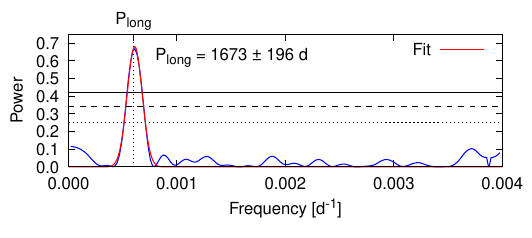}
  \caption{
    GLS periodogram of the synthetic data, containing $P_1$, $P_2$, and $P_{\rm long}$, and averaged over 129~d. Despite the averaging, a period at $P_{\rm long} = 1640 \pm 183$ is very significantly detected.
  }
  \label{fig:synth-avg-ls}
\end{figure}

Finally, we average the synthetic data of the model containing all three periods over time bins of width 129~days. This is the same approach as we applied to the observational OVRO data in Sect.~\ref{sec:filter} and Fig.~\ref{fig:OVRO-avg}. The GLS periodogram of the averaged synthetic data is plotted in Fig.~\ref{fig:synth-avg-ls}. In this plot there is only one significant peak, and the period is determined as $1640 \pm 183$~d after fitting a one-peaked Gaussian, which is identified to be the long-term period $P_{\rm long}$. The power of the peak in the synthetic data (0.7) is a little higher than for the observational data (0.5), which is expected from the absence of noise in the synthetic data. This same behavior of the observational and synthetic data of retaining $P_{\rm long}$ after averaging out the short-term periods $P_1$ and $P_2$ (c.f.\ Fig.~\ref{fig:OVRO-avg}\,b) further corroborates the scenario in which $P_{\rm long}$ is indeed an intrinsic (physical) period of the system. It does, however, not rule out the possibility that also $P_2$ (along with $P_1$) is a physically real period in the system (as shown above).

\section{Conclusions} \label{sec:conclusions}

In this paper, we  analyze archival radio data for the \gammaray{} emitting  binary \lsi{}. We  compare the results of timing-analysis of these observational data to those obtained from purely synthetic data. In all aspects of our analysis, we  avoided assuming any physical scenario for the processes responsible for the periodic signals from this source. Our aim was rather to identify the intrinsic (physical) periods of this system based on the analysis of the variability of its radio emission alone. Our main conclusions are listed below. 

\begin{enumerate}
\item{
  The periods $P_1 = 26.5$~d (orbital) and $P_2 = 26.9$~d (quasi-orbital) were only significantly detected when analyzing a long-enough light curve. This is expected from the increased frequency resolution of a longer time-series in Fourier space.
}
\item{
  A two-peaked profile can be detected with data that span less than one cycle of the $P_{\rm long} = 4.6$~year long-term modulation if that dataset includes the long-term minimum. This strongly indicates that $P_2$ is not a beat of $P_1$ and $P_{\rm long}$ and, instead, its origin could be rooted in a physical process modulated with $P_2$.
}
\item{
  Analysis of synthetic data, based on the sampling of the OVRO dataset, suggests that, in fact,  all three periods, $P_1$, $P_2$, and~$P_{\rm long}$, are physically real intrinsic periods of the system. The  superposition of three sine waves oscillating at these periods reproduces the observed data surprisingly well. The remaining noise can be attributed to stochastic processes in the source itself.
}
\item{
  The physical reality of all three periods suggests a scenario resembling the framework described at the end of Sect.~\ref{physsc}. In particular, it would call for a physical process that is periodic with $P_{\rm long}$, which would indeed occur in \lsi{} (plausibly, but not necessarily, related to the Be disk). This process physically affects the emitting region (e.g., via accretion, jet or pulsar wind-star outflow interactions, jet precession, etc.), as its long-term modulation induces, orbit by orbit, a small delay in the orbital phase of the non-thermal emission maximum. The latter process thus mimics the effect of a beating between $P_1$ and $P_{\rm long}$, but $P_2$ is intrinsic of the emitting region, not a mathematical artifact resulting from the evolution of two unrelated structures with periods, $P_1$ and $P_{\rm long}$.
}
\item{
    A period with a value of $P_3 = 26.3$~d (as reported by \citealt{Zhang2024}) has been significantly detected when revisiting the OVRO data of \citet{Jaron2024}. However, this period also significantly appears in the periodograms of all synthetic data, without any need to explicitly including it in the model. We thus conclude that this is not an intrinsic period of the system, but the result of a beating between the intrinsic periods of the system.
}
\end{enumerate}

We have shown that a purely phenomenological study of the variability of the emission of \lsi{} can shed light on important properties of the physical structures and processes underlying that emission. Therefore, continued monitoring of the intriguing binary system at radio and other wavelengths is highly encouraged.

\begin{acknowledgements}
  We thank Gunther Witzel of the MPIfR for reading the manuscript and for useful feedback.
  This research has made use of data from the OVRO 40-m monitoring program \citep{Richards2011}, supported by private funding from the California Institute of Technology and the Max Planck Institute for Radio Astronomy, and by NASA grants NNX08AW31G, NNX11A043G, and NNX14AQ89G and NSF grants AST-0808050 and AST-1109911.
  FJ acknowledges funding by the Austrian Science Fund (FWF) [P35920]. VB-R acknowledges financial support from the Spanish Ministry of Science and Innovation under grant PID2022-136828NB-C41/AEI/10.13039/501100011033/ERDF/EU and through the Mar\'ia de Maeztu award to the ICCUB (CEX2024-001451-M). VB-R is Correspondent Researcher of CONICET, Argentina, at the IAR.   
\end{acknowledgements}

\bibliographystyle{aa}
\bibliography{lit.bib}

\begin{appendix}

In Sect.~\ref{sec:results}, we fit the observational OVRO data with the superposition of sine functions of the form given in Eq.~(\ref{eq:synth}). The parameters resulting from performing this fit are listed in Table~\ref{tab:fitresults}.

\begin{table}[]
  \section{Parameters from synthetic data fit}
    \centering
    \caption{
        Parameters resulting from fitting the OVRO 15~GHz data with a function of the form given in Eq.~(\ref{eq:synth}), for each of the period combinations. The indices of the parameters given here correspond to the usual definitions of the periods of \lsi{}.
    }
    \label{tab:fitresults}
    \begin{tabular}{cr}
    \hline
    \hline
    Model: & $P_1$ + $P_2$\\
    \hline
    $P_1$ & $26.493 \pm 0.001$~d\\
    $P_2$ & $26.913 \pm 0.001$~d\\
    $a_1$ & $22.2 \pm 0.2$\,mJy\\
    $a_2$ & $15.2 \pm 0.2$\,mJy\\
    $\phi_1$ & $2.36 \pm 0.37$\,rad\\
    $\phi_2$ & $-5.53 \pm 0.56$\,rad\\
    $c$ & $38.0 \pm 0.1$\,mJy\\
    \hline
    \hline
    Model: & $P_1$ + $P_{\rm long}$\\
    \hline
    $P_1$ & $26.492 \pm 0.001$~d\\
    $P_{\rm long}$ & $1615 \pm 10$~d\\
    $a_1$ & $21.1 \pm 0.2$\,mJy\\
    $a_{\rm long}$ & $6.6 \pm 0.2$\,mJy\\
    $\phi_1$ & $3.04 \pm 0.39$\,rad\\
    $\phi_{\rm long}$ & $4.91 \pm 1.46$\,rad\\
    $c$ & $37.1 \pm 0.1$\,mJy\\
    \hline
    \hline
    Model: & $P_1$ + $P_2$ + $P_{\rm long}$\\
    \hline
    $P_1$ & $26.492 \pm 0.001$~d\\
    $P_2$ & $26.917 \pm 0.001$~d\\
    $P_{\rm long}$ & $1642 \pm 10$~d\\
    $a_1$ & $23.2 \pm 0.2$\,mJy\\
    $a_2$ & $15.3 \pm 0.2$\,mJy\\
    $a_{\rm long}$ & $6.7 \pm 0.2$\,mJy\\
    $\phi_1$ & $2.66 \pm 0.36$\,rad\\
    $\phi_2$ & $-7.41 \pm 0.56$\,rad\\
    $\phi_{\rm long}$ & $0.98 \pm 1.32$\,rad\\
    $c$ & $38.3 \pm 0.1$\,mJy\\
    \hline
    \hline
    \end{tabular}
\end{table}

\end{appendix}

\end{document}